\documentclass[12pt]{extarticle}
\usepackage{tikz}
\usepackage{listings}
\usepackage{xcolor}
\usepackage{amsmath}
\usepackage{amssymb}
\usepackage[export]{adjustbox}
\usepackage{makecell}
\usepackage{adjustbox}
\usepackage{float}
\usepackage{gensymb}
\usepackage[margin=1in]{geometry}
\usepackage{graphicx}
\usepackage{titlesec}
\usepackage[final]{changes}
\usepackage[
  backend=biber,
  style=numeric-comp,
  sorting=none,
  maxnames=2,
  minnames=1,
  url=false,
  doi=true,
  isbn=false
]{biblatex}
\title{The role of absorption in three-dimensional electron diffraction dynamical structure refinement}
\usepackage{authblk}

\title{The role of absorption in three-dimensional electron diffraction dynamical structure refinement}

\author[1]{Benjamin Colmey}
\author[1,2]{Tiarnan A.S. Doherty}
\author[2]{Shreshth A. Malik}
\author[1]{Paul A. Midgley}

\affil[1]{Department of Materials Science and Metallurgy, University of Cambridge, 27 Charles Babbage Rd, Cambridge, CB3 0FS, United Kingdom}
\affil[2]{OATML, Department of Computer Science, University of Oxford, Wolfson Building, Parks Rd, Oxford, OX1 3QG, United Kingdom}

\date{}

\begin{document}

\maketitle

\begin{abstract}

The role of absorption in 3D electron diffraction is established through analytical theory, simulation, and dynamical refinement. A two-beam expression for the absorbed integrated intensity \added{in centrosymmetric crystals} is derived, showing that for $t/\xi_g \ll 1$ reflections follow a uniform exponential decay set by the mean absorptive potential $U_0'$. 
\replaced{Many-beam simulations of both centrosymmetric and non-centrosymmetric crystals reveal  additional reflection-specific anomalous absorption beyond the uniform attenuation set by $U_0'$. Neglecting these effects in dynamical refinement of integrated intensities incurs an error that increases approximately linearly with thickness, with this error becoming more severe near zone axes.}{ Many-beam simulations demonstrate that neglecting absorption in dynamical refinement of integrated intensities incurs a residual that increases linearly with thickness, and diverges near zone axes.} Dynamical refinements were performed on CsPbBr$_3$, quartz, and borane, with the inclusion of absorption yielding an improvement  in $R_{\mathrm{obs}}$ from $6.4$ to $5.3$ \% for CsPbBr$_3$ and negligible improvements for quartz and borane. \added{Anomalous} absorption \replaced{may therefore be ignored}{ is therefore deemed negligible} for routine refinement of integrated intensities except in high-$Z$ materials at thicknesses approaching $\xi_g $.

\end{abstract}

\section{Introduction}

Electron diffraction (ED) has become an established tool for determining the structures of nanocrystalline materials that are inaccessible to conventional X-ray crystallography. Developments such as precession electron diffraction (PED) \cite{Vincent1994} and continuous-rotation 3D ED have expanded its applicability, enabling routine structure determination across inorganic, hybrid, and molecular crystals \cite{Gemmi2019}.

Building on this foundation, dynamical refinement against 3D~ED data \cite{Palatinus:ra5008,Palatinus:td5023}, as implemented in e.g. \textit{JANA2020} \cite{Petříček}, has enabled a continuing series of breakthroughs, including hydrogen localization \cite{Palatinus2017}, multipolar refinements \cite{Olech2024} and ionisation-state analysis via $\kappa$-refinement \cite{Suresh2024}. Alternative formulations of dynamical refinement continue to be developed \cite{malik2025hybrid}, extending the range and precision of these methods.

Refinement is the process by which a structural model is iteratively adjusted until convergence is achieved between simulated and observed diffraction intensities. Its accuracy is governed by the extent to which the scattering simulation reproduces the true physical interaction. The kinematical approach offers a simplified model, while dynamical simulation captures the full complexity of multiple scattering and can achieve far higher physical fidelity.

Despite these advances, inelastic scattering has generally been neglected in 3D ED refinement \cite{Saha, Palatinus2013, thomas2024parameterized}. Such processes are known to alter diffracted intensities in ways not captured by purely elastic models \cite{Eggeman2013PED}. Their omission thus limits the completeness of current refinements, potentially introducing artifacts into solved structures.

In high-energy ED, where accelerating voltages typically exceed 50 keV, incident electrons can exchange energy with the material through three inelastic mechanisms: thermal-diffuse scattering (TDS), plasmon excitations, and core-loss ionisation. Historically, the term \textit{absorption} has been used to describe the effects arising from TDS. In reality, no true absorption occurs; rather, these processes lead to a loss of elastic intensity and a redistribution of diffracted intensity among beams \cite{humphreys1968absorption}, in contrast to the genuine absorption of photons in X-ray diffraction \cite{Albrecht1939Absorption}. Here, it is worth noting that although many kinematical 3D ED refinements make use of X-ray software and absorption corrections \cite{vanGenderen2016, simoncic2023electron, leung2024formulation}, these corrections model a fundamentally different phenomenon.

While the electron-specific absorption formalism has previously been incorporated in convergent-beam electron diffraction (CBED) refinements \cite{Bird1992_CBED,Zuo1991_CBEDRefinement,Midgley1995_EFCBED,Zuo1999_Cu2O,Nakashima:pl5049} its use in 3D ED remains limited. Beyond CBED, absorption has been addressed through simulations of precession data using a constant absorptive potential, where comparisons with elastic-only calculations revealed only minor differences, but no systematic investigation of the impact of absorption on refinement was undertaken \cite{Palatinus2013}.

In CBED, a stationary crystal is illuminated by a convergent probe, producing Bragg discs that encode the full two-dimensional region surrounding a reciprocal-lattice vector. In 3D ED and PED, crystal rotation or beam precession integrates intensities over linear or circular trajectories, effectively averaging orientation-specific dynamical contrast. The consequence is a set of integrated intensities that more closely match the kinematical approximation \cite{Blackman1939}, making them well suited for structure solution techniques originally devised for X-ray data. 

These considerations raise the central question: under what conditions does absorption materially influence structure refinement, and when can it safely be neglected? To address this, Bloch-wave simulations were performed with and without absorption across multiple materials and orientations, revealing its role on integrated intensities. Dynamical refinements including absorption were subsequently performed on experimental data, establishing the practical limits of the elastic-only approximation.

 \section{Methods}
\subsection{Thermal Effects in Quantitative Electron Diffraction}

The present analysis builds upon the dynamical refinement method reported by Malik \emph{et al.}~\cite{malik2025hybrid}, which implements a differentiable Bloch-wave framework. Unlike the kinematical approximation, which assumes single scattering, Bloch-wave methods explicitly account for coupling between diffracted beams. 

Thermal vibrations in crystals give rise to diffuse scattering that removes electrons from the coherent diffracted beams, thereby producing an attenuation of diffracted intensities \cite{Hashimoto1962}. These effects can be decomposed into TDS, arising from random atomic displacements and phonon scattering \cite{Wang1992, Eggeman2013}. Both processes produce sub-eV energy losses that fall within the `zero-loss' peak, rendering most commercially available energy filters ineffective at removing absorption effects and requiring treatment through cooling or explicit modeling \cite{Nakashima:pl5049}. \added{Beyond thermal effects, additional inelastic scattering processes, sometimes also grouped under the term absorption, such as plasmon scattering and core-loss ionisation can, in principle, be removed through energy filtering, although the extent to which this improves diffraction refinement remains unclear} \cite{Jansen2004,Eggeman2013PED,Yang2022,Kim2010,Gemmi2013,palatinus2015structure,Latychevskaia2019}.

While the original Bloch wave formalism developed by Bethe \cite{Bethe1928} treats only elastic scattering, \replaced{the effects of TDS}{ absorption} can be included through the complex Fourier coefficient of the potential \cite{Yoshioka1957}. For a reciprocal lattice vector $\mathbf g$, this coefficient \replaced{can be approximated }{ can be written}  as \cite{bird1990absorptive}
\begin{align}
U^{\text{tot}}_\mathbf{g} = U_\mathbf{g} + iU'_\mathbf{g} =
-\frac{h^2}{2m_0 \Omega}
\sum_\kappa e^{-2 i \pi (\mathbf{g}\cdot\mathbf{r}_\kappa)}
\left[f_\kappa(s) + i f'_\kappa(s, B_\kappa)\right]
e^{-B_\kappa s^2}
\end{align}

where $s = |\mathbf g|/2$, $\mathbf{r_\kappa}$ is the position of atom $\kappa$ in the unit cell, $m_0$, $h$, and $\Omega$ denote the electron rest mass, Planck's constant, and the unit-cell volume, respectively. Here $f_\kappa(s)$ is the elastic scattering factor and $f'_\kappa(s,B_\kappa)$ its absorptive counterpart, \replaced{ calculated within the Einstein model approximation using} {expressed in terms of $s$ and} the isotropic Debye--Waller factor $B_\kappa = 8\pi^2 \langle u_\kappa^2 \rangle$, where $\langle u_\kappa^2 \rangle$ is the mean-squared vibrational displacement.

This displacement depends on the stiffness of the interatomic potential. Materials with high Debye temperatures (\(\Theta_D\)) maintain relatively small \(\langle u^2\rangle\)  when compared to materials with low \(\Theta_D\) under the same conditions. Cooling reduces \(\langle u^2\rangle\) and correspondingly diminishes the magnitude of the absorptive effects \cite{humphreys1968absorption}.

Despite its quantitative improvements over the elastic-only model, this formulation remains approximate. A principal limitation is the assumption that absorbed electrons are removed rather than redistributed into diffuse background intensity \cite{humphreys1968absorption}. Although background subtraction mitigates the effect, it cannot fully separate diffuse inelastic contributions from the elastic signal. \added{Analytical treatments of TDS suggest that thermally scattered
electrons, while centred on Bragg discs, are distributed over far larger
scattering angles than elastically scattered electrons} 
\cite{Wang1995}. \added{As an example, for InP
at 250 keV and 300 K, only \(4.2\%\) of TDS electrons are expected to
fall within a \(10~\mathrm{mrad}\) semi-angle} \cite{Jordan1991EnergyFilteringLACBED}; \added{for a conservatively large Bragg-disc radius of \(2.5~\mathrm{mrad}\), the resulting influence on the measured Bragg intensities is therefore expected to be small.}

Another important approximation concerns the use of isotropic atomic vibrations, represented by a single parameter \( U_{\mathrm{iso}} \), or calculated as $U_{\mathrm{iso}} = \frac{1}{3}(U_{11} + U_{22} + U_{33})$ when anisotropic displacement parameters  are available. While anisotropic forms of $f'_\kappa$ exist, they are considerably more demanding computationally, and the errors introduced in $f'_\kappa$ by using isotropic Debye–Waller factors are typically on the order of 1–5 \% for low- and medium-order reflections \cite{Peng1997}. 

A further limitation arises from the use of the Einstein model, which assumes independent oscillators and fails to describe systems with strong phonon correlations. A more complete description is provided by the frozen phonon method \cite{Wang1995,Allen2015_inelastic}, which conserves total scattering intensity and recovers the diffuse background by averaging over many thermal configurations. While frozen phonon can be incorporated into the Bloch-wave framework  \cite{yamazaki2013,Mendis2025}, it typically requires averaging over dozens of thermal configurations, imposing a computational burden that restricts its practical use in refinement.

\subsection{Two-Beam Absorption Formalism }

The two-beam approximation represents the simplest extension beyond the kinematical model, coupling the incident and a single diffracted beam while neglecting higher-order interactions. Although it fails to describe the complex coupling present in strongly diffracting systems or near high-symmetry zone axes, it is often valuable for analytical derivations and in theoretical interpretation.

The two-beam approximation gives analytical expressions for the diffracted-beam intensity under absorption \added{in centrosymmetric crystals} \cite{Hashimoto1962}:
\begin{equation}
|\Psi_{\mathbf{g}}|_\mathrm{abs}^2 = e^{(-2 \kappa_0 t)} \frac{(\cosh(2 \Delta \kappa t) - \cos(2 \pi \Delta k t))}{2(1 + x^2)}  
\label{eq:two_beam_absorption}
\end{equation}
%\begin{equation}
% |\Psi_{\mathbf{g}}|^2 = 1- |\Psi_0|^2
% \end{equation}
with the following definitions:
\begin{align*}
\kappa_0 &= \frac{\pi}{\xi_0'} \quad\quad
\Delta k = \frac{\sqrt{1 + x^2}}{\xi_{\mathbf{g}}} \quad\quad
\Delta \kappa = \frac{\pi}{\xi_{\mathbf{g}}' \sqrt{1 + x^2}} \\[0.5em]
\xi_0' &= \frac{K}{U_0'} \quad\quad
\xi_{\mathbf{g}}' = \frac{K}{U_{\mathbf{g}}'} \quad\quad
\xi_{\mathbf{g}} = \frac{K}{|U_{\mathbf{g}}|} \quad\quad
x = \xi_{\mathbf{g}} S_{\mathbf{g}}
\end{align*}

%given by. mention small beam approximation cos(theta) goes to 1. 

\added{
Here, \(K = |\mathbf K| = 1/\lambda\) is the magnitude of the incident electron wavevector \(\mathbf K\), corresponding to the radius of the Ewald sphere. The excitation error $S_g$ is defined as }\cite{Spence2020} \replaced{
\[
S_{\mathbf g}
=
\frac{K^2-|\mathbf K+\mathbf g|^2}{2K},
\]
which may be interpreted as the distance of a reciprocal-lattice point from the Ewald sphere surface.
}{
Here, $S_{\mathbf{g}}$ is the distance of a reciprocal lattice vector from the Ewald sphere surface and \(K\) is the radius of the Ewald sphere, given by \(K = 1 / \lambda\).
}

%Here, \replaced{\(K\) is the radius of the Ewald sphere, given by \(K = 1 / \lambda\) and $S_{\mathbf g}=\frac{K^2-|\mathbf{K}+\mathbf g|^2}{2K}$, can be interpreted as the distance of a reciprocal lattice vector from the surface of the Ewald sphere.} { \quad Here, $S_{\mathbf{g}}$ is the distance of a reciprocal lattice vector from the Ewald sphere surface and \(K\) is the radius of the Ewald sphere, given by \(K = 1 / \lambda\)} . 
The quantity \(\xi'_0\) determines the mean absorption, while \(\xi'_{\mathbf{g}}\) governs the strength of anomalous absorption; \cite{Hashimoto1962} \replaced{describe $\xi'_g$ as the absorption distance.}{and may be regarded as an absorption distance, by analogy with the extinction distance \(\xi_{\mathbf{g}}\)}.

The mean absorptive potential \(U'_0\), is defined as \cite{rez}
\begin{equation}  
U'_0 = \frac{\gamma}{\pi} \, \frac{\sum_{i} n_i f'_i(0)}{\Omega},
\end{equation}
where \(\gamma\) is the relativistic correction factor, \(n_i\) are the site occupancies, \(f'_i(0)\) are the absorptive scattering factors at \(\mathbf{g}=0\) and the sum is over all atoms in the unit cell. In the absence of absorption, the two-beam diffracted intensity simplifies to \cite{hirsch1965electron}:  %Eq. (8.23):  %, and \(\Omega\) is the unit-cell volume.
\begin{equation}
|\Psi_{\mathbf{g}}|_\mathrm{no\, abs}^2 =
\frac{\sin^2\!\left(\pi t\,\Delta k\right)}
     {(1 + x^2)},
\label{eq:two_beam_no_abs_x}
\end{equation}
which can be rewritten as
\begin{equation}
|\Psi_{\mathbf{g}}|_\mathrm{no\, abs}^2 =
\frac{1 - \cos(2\pi t\,\Delta k)}{2(1 + x^2)},
\label{eq:two_beam_no_abs_cos}
\end{equation}

In Eq.~\ref{eq:two_beam_absorption}, absorption enters through a global damping term \(e^{-2\kappa_0 t}\) and a reflection specific term \(\cosh(2\Delta\kappa t)\). When \(U_0' = U_{\mathbf{g}}' = 0\), both reduce to unity, and Eq.~\ref{eq:two_beam_absorption} reverts to the elastic form of Eq.~\ref{eq:two_beam_no_abs_cos}. While Eqs.~\ref{eq:two_beam_absorption} and \ref{eq:two_beam_no_abs_cos} are given for diffraction at a specific orientation $S_g$, in 3D~ED the quantity of interest is the integrated rocking-curve intensity. Starting from the non-absorbed two-beam expression (Eq.~\ref{eq:two_beam_no_abs_cos}), the relevant quantity is

\begin{equation}
I_{\mathrm{no\,abs}}^{\mathrm{int}}(t)
= \int_{-\infty}^{\infty} |\Psi_{\mathbf{g}}|_{\mathrm{no\,abs}}^{2}\, dx , \qquad x = \xi_{\mathbf{g}} S_{\mathbf{g}} 
\label{eq:I_noabs_integrated_def}
\end{equation}

\added{Following Blackman's treatment of two-beam integrated intensities}
\cite{Blackman1939}\added{, we introduce \(A=\pi t/\xi_{\mathbf g}\), giving}
\cite{own2005precession}
\replaced{}{
This integral has the well-known analytical form first derived by Blackman.
Introducing \(A=\pi t/\xi_{\mathbf g}\), one obtains
}

\begin{equation}
I_{\mathrm{no\,abs}}^{\mathrm{int}}(t) 
\added{=}
\int_{-\infty}^{\infty}
\frac{
\sin^{2}\!\left(A\sqrt{1+x^{2}}\right)
}{
1+x^{2}
}\,dx
\added{=}
\added{\pi}
\int_{0}^{A}J_{0}(2u)\,du .
\label{eq:I_noabs_blackman}
\end{equation}

where \(J_{0}\) is the zeroth-order Bessel function of the first kind.

As derived in Supplementary Section~S1, carrying out the corresponding
integral for the absorbed case in the weak-absorption limit
(\(t/\xi_{\mathbf g}\ll1\)) yields

\begin{equation}
\frac{I_{\mathrm{abs}}^{\mathrm{int}}(t)}
     {I_{\mathrm{no\,abs}}^{\mathrm{int}}(t)}
\approx
e^{-2\kappa_0 t}
\left[
1
+
\frac{\pi t\,\xi_g}
     {2\xi_g'^2}
\left(
1
-
\frac{\pi^2 t^2}
     {3\xi_g^2}
{
+\frac{\pi^4 t^4}{20\xi_g^4}
}
\right)^{-1}
\right].
\label{eq:Iabs_int_clean}
\end{equation}

%As demonstrated in Supplementary Section S1, the analytical form given here produces values in good agreement with those obtained from direct numerical integration. 
\added{For centrosymmetric crystals,} absorption therefore modifies the integrated two-beam intensity in two distinct ways. The global factor \(e^{-2\kappa_0 t}\) applies a uniform thickness-dependent attenuation, given by \(U_0'\). The second factor represents the anomalous
contribution arising from the reflection-specific
\(U_{\mathbf{g}}'\).  At large thickness, this anomalous term causes reflections to diverge from the uniform decay. Reflections with large $U_{\mathbf{g}}'$ (small $\xi_{\mathbf{g}}'$) show the largest departures, since the anomalous term scales approximately as $\xi_{\mathbf{g}} / \xi_{\mathbf{g}}'^2$.

\added{For non-centrosymmetric crystals, additional first-order absorptive contributions may also arise from phase differences between the elastic and absorptive structure factors} \cite{bird1990noncentro}.  \added{These effects are discussed in Supplementary Section~S2, where additional simulations indicate that the resulting additional non-centrosymmetric contributions remain comparatively small.}

%Thus, reflections with large $U_{\mathbf{g}}'$ display less overall attenuation. 

To obtain experimentally meaningful quantities, the two-beam integrated expressions \added{in Eqs.~\ref{eq:I_noabs_integrated_def}--\ref{eq:Iabs_int_clean}} must include an additional Lorentz correction \replaced {of the form $I_{\mathrm{2\text{-}beam}}^{\mathrm{int}} \times L$, where $L$ accounts} { This correction accounts} for the angular dwell time of a reflection near its Bragg condition, which is inversely proportional to the velocity of Bragg-crossing during rotation \cite{McIntyre:du0066,Zhang}.  

In a continuous rotation experiment, this velocity is set by the component of the
reciprocal-lattice vector perpendicular to the rotation axis \cite{Holmes}. If $\phi$
denotes the angle between $\mathbf{g}$ and the rotation axis, the
intersection speed is proportional to $|\mathbf{g}|\sin\phi$.  The correction therefore takes the form $
L = {1}/{|\mathbf{g}|\sin\phi}$,
which expresses the fact that reflections near the rotation axis remain near their Bragg condition for longer and contribute more strongly to the observed experimental intensity.

\section{Results}
\subsection{Comparison of Two- and Many-Beam Models of Absorption}
The material caesium lead bromide (\(\mathrm{CsPbBr_3}\)) was examined using the structure reported in ~\cite{Suresh2024}, with the corresponding structural parameters listed in Table~\ref{tab:material_parameters}. This compound was selected because it contains atoms of large atomic number and correspondingly strong elastic and absorptive scattering factors $f$ and $f'$, making absorption especially significant.

Two-beam and many-beam integrated intensities were simulated for \(\mathrm{CsPbBr_3}\) at 200 keV using the first experimentally observed orientation, \added {[\textit{uvw}] = [-0.43, 1.00, 0.93], taken from the 3D ED dataset reported in (Suresh et al. 2024) and determined during data reduction using PETS2} \cite{palatinus2011pets}.  This orientation defines the centre of a {virtual frame}, as described in \cite{klar_accurate_2023}, with each virtual frame corresponding to a small angular range in \(\alpha\) around the central orientation, where \(\alpha\) is the rotation axis of the goniometer. In this study, all orientations correspond to experimentally observed ones,  while specific angular positions within a virtual frame are referred to as tilts.

The two-beam integrated intensities for this orientation were computed by numerically integrating Eqs.~\eqref{eq:two_beam_absorption} and \eqref{eq:two_beam_no_abs_x} with an additional Lorentz correction. In the many-beam case, calculations were performed using the full Bloch-wave code described in \cite{malik2025hybrid}, with the parametrized form of the absorptive scattering factors \(f'\) adapted from \cite{thomas2024parameterized}. Here, no Lorentz correction is required because the rotation geometry is already encoded through explicit tilt sampling.

A \(\pm 1.5^{\circ}\) sweep around the experimental orientation was simulated with 60 uniformly spaced tilt samples, and the corresponding rocking curves were summed to obtain the integrated intensities. Unless otherwise stated, 60 tilt samples per orientation were used throughout. 

Simulations were performed for thicknesses between 10--250 nm. This lower limit was chosen as the width of rocking curves scale inversely with thickness in the kinematic limit \cite{hirsch1965electron}, and therefore at very small thicknesses (where the kinematic approximation is valid), the curves may exceed the angular range defined by the virtual frame. As a result, part of the diffracted intensity may fall outside the sampling window, making these reflections not fully-integrated.

\begin{figure}[H]    
    
    \includegraphics[width=1\linewidth]{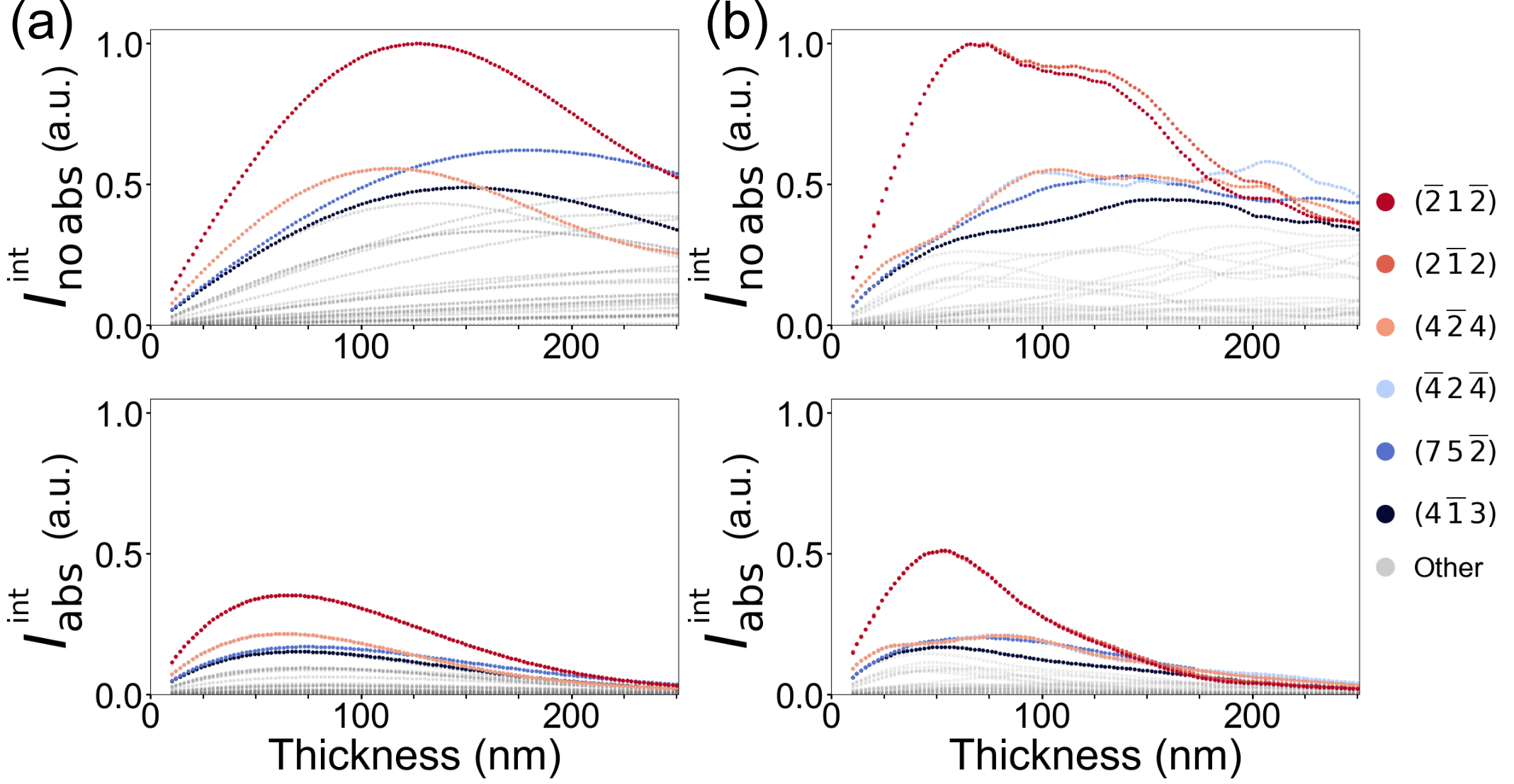}
    \caption[]{Integrated intensity vs thickness for {CsPbBr$_3$}, 200\,keV, \([uvw]=[-0.43,  1.00,  0.93]\), 
     without (top) and with absorption (bottom), for (a) two-beam, (b) many-beam model. The six most intense reflections are highlighted in colour, with all others shown in grey for clarity.}
\label{fig:doublepend}
\end{figure}

The resulting two-beam and many-beam integrated intensities, with and without absorption, are shown in Fig.~\ref{fig:doublepend}.

While the two-beam curves show smooth Pendellösung-type  oscillations, the many-beam case develops the expected irregularities from multiple strongly coupled reflections. In the two-beam model, pairs such as $(\bar{2}\,1\,\bar{2})/(2\,\bar{1}\,2)$ and  $(4\,\bar{2}\,4)/(\bar{4}\,2\,\bar{4})$  maintain identical intensities over the full thickness range. In the many-beam calculation, however, these pairs begin to diverge at thicknesses approaching $100$~nm, reflecting the redistribution of intensity through additional dynamical pathways.

\begin{figure}[H]
    \includegraphics[width=\linewidth]{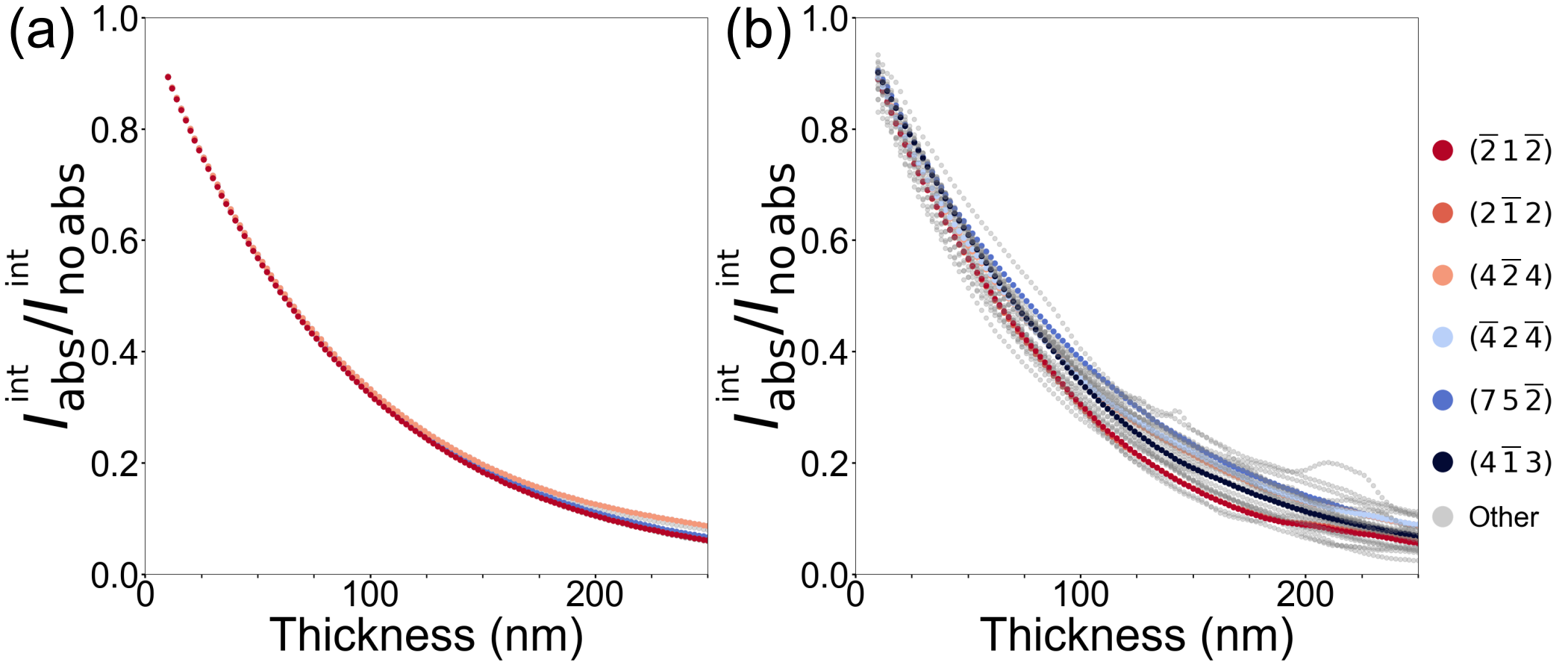}
    \caption[Two-beam simulations for CsPbBr$_3$]{Ratio of integrated intensity
    ($I^{\mathrm{int}}_{\mathrm{abs}}/I^{\mathrm{int}}_{\mathrm{no\,abs}}$) as a function of thickness 
    for $\mathrm{CsPbBr_3}$ with \([uvw]=[-0.43,\,1.00,\,0.93]\).  
    (a) Two-beam model; (b) many-beam model.  
    A decaying exponential was fitted to each $\textit{hkl}$ curve, yielding absorption parameters 
    $\bar{\lambda}=89.9\,\mathrm{nm}$, $\sigma_{\lambda}=2.0\,\mathrm{nm}$ (two-beam) and  
    $\bar{\lambda}=93.2\,\mathrm{nm}$, $\sigma_{\lambda}=10.1\,\mathrm{nm}$ (many-beam).}
    \label{fig:two-beam}
\end{figure}

 In both models, absorption clearly attenuates intensities, but this plot alone does not reveal the reflection-specific decay rates or how the models compare. To address this comparison, Fig.~\ref{fig:two-beam} shows the ratio of absorbed to non-absorbed integrated intensities ($I_{\mathrm{abs}}^{\mathrm{int}}/I^{\mathrm{int}}_{\mathrm{no\,abs}}$) as a function of thickness for the two and many-beam models.

In the two-beam case (Fig.~\ref{fig:two-beam}(a)), fitting an exponential decay to each individual \(hkl\) yields a mean absorption length of \(\bar{\lambda}=89.9~\text{nm}\) with a standard deviation of \(\sigma_{\lambda}=2.0~\text{nm}\). From the two-beam absorption formalism, the decay constant is determined by the global damping term, with \(\lambda_{\mathrm{2\text{-}beam}} = K / (2\pi U_0')\). Using \(K = 39.87~\text{\AA}^{-1}\) at 200~keV and \(U_0' = 0.0072~\text{\AA}^{-2}\) for CsPbBr\(_3\), the predicted absorption length is \(\lambda_{\mathrm{2\text{-}beam}} = 88.4~\text{nm}\), in excellent agreement with the fitted value.

As expected, the two-beam model shows little reflection-dependent variation in the absorption ratios. $(4\,\bar{2}\,4)$ and $(\bar{4}\,2\,\bar{4})$ \replaced{begin to depart from the otherwise nearly uniform attenuation trend}{ begin to deviate} for $t >150$~nm ($(4\,\bar{2}\,4)$ is not visible because it lies directly beneath $(\bar{4}\,2\,\bar{4})$). These reflections have the largest values of $U_{\mathbf{g}}$ and $U_{\mathbf{g}}'$ among the observed set, with $U_{\mathbf{g}} = 0.013~\text{\AA}^{-2}$, $U_{\mathbf{g}}' = 0.003~\text{\AA}^{-2}$, giving  $\xi_{\mathbf{g}} \approx 300$~nm and $\xi_{\mathbf{g}}' \approx 1400$~nm. Eq.~\ref{eq:Iabs_int_clean} predicts that the anomalous contribution increases when $t$ approaches $\xi_{\mathbf{g}}$, in agreement with the observed behaviour. 
\added{To further illustrate the deviation from uniform attenuation in the two-beam model as \(t\) approaches \(\xi_g\), an extended version of Fig.~\ref{fig:two-beam}(a) covering 0--350~nm is shown in Supplementary Fig.~S1 (a).}

By contrast, the many-beam case (Fig.~\ref{fig:two-beam}(b)) exhibits a broader distribution of decay constants, with \( \bar{\lambda}=93.2~\text{nm} \) with \(\sigma_{\lambda}=10.1~\text{nm}\), indicative of the effects of anomalous, reflection-specific absorption.

Beyond integrated intensities, rocking-curve analysis provides further insight into how absorption modifies scattering pathways. Many-beam rocking curves for all reflections considered in the above calculation at  $t$ = 200 nm are shown in Supplementary Fig. S5. While not representative of the original sample thickness observed in \cite{Suresh2024}, 200 nm serves as an illustrative high-thickness example, where absorption is especially pronounced.

Supplementary Fig. S5 shows that absorption can modify the tilt-dependent intensity profile (i.e. rocking curve) in ways that vary markedly across reflections: for some, the effect is nearly uniform across all tilts, while for others it is strongly tilt dependent. Despite these local tilt variations, the corresponding integrated intensities exhibit far smaller differences between absorptive and elastic simulations. This contrast helps explain why absorption plays a significant role in CBED analysis, where rocking-curve shapes are fitted directly \cite{Zuo1991_CBEDRefinement}, but a reduced one in 3D~ED, where integration suppresses tilt-specific behaviour.

\subsection{Implications of Neglecting Absorption in Dynamical Refinement}

The discrepancy between elastic and absorptive simulations can be further quantified using a residual. A standard choice is $R_1$, defined as
\[
R_1 =
\frac{
\sum_i \left|\sqrt{\vphantom{I^{I^{I}}} I_i^{\mathrm{exp}}}
            - \sqrt{\vphantom{I^{I^{I}}} I_i^{\mathrm{calc}}}\right|
}{
\sum_i \sqrt{\vphantom{I^{I^{I}}} I_i^{\mathrm{exp}}}
}.
\]

where $I_i^{\mathrm{exp}}$ denotes the experimentally observed intensity of reflection $i$, and $I_i^{\mathrm{calc}}$ the corresponding simulated intensity. Because $R_1$ does not weight reflections by their experimental uncertainties, it is particularly sensitive to noisy, low-intensity reflections. Consequently, practical refinements typically report \(R_{\mathrm{obs}}\), defined as \(R_1\) evaluated only over reflections satisfying \(I^{\mathrm{exp}}_i/\sigma_i > 3\), where \(\sigma_i\) is the estimated measurement error \cite{Palatinus2019}. Another alternative is the weighted form \(wR_{\mathrm{all}}\), which considers all intensities and weights the contribution of each \(I_i\) according to its uncertainty \(\sigma_i\) and its magnitude.
To estimate the influence of absorption,  \(R_1\) can be adapted to measure the deviation between absorptive and purely elastic simulations.
\begin{equation}
R_1(t)
=
\frac{
\sum_i \left|\sqrt{I_{i}^{\mathrm{int,abs}}(t)} - \sqrt{I_{i}^{\mathrm{int,no \,abs}}}\right|
}{
\sum_i \sqrt{I_{i}^{\mathrm{int,no\, abs}}}
},
\label{eq:rabs}
\end{equation}
\added{Here, a global scale factor corresponding to the mean absorptive attenuation has been omitted for simplicity, such that only reflection-dependent deviations contribute to the residual; the full derivation is given in Supplementary Section S3.} If the ratio of integrated intensities \(I_{i}^{\mathrm{int,abs}}/ I_{i}^{\mathrm{int,no\,abs}}\) is approximated as an exponential decay with mean decay constant $\bar{\lambda}$ and a Gaussian spread of $\lambda_i$, the resulting thickness–dependence of the residual can be estimated in closed form:
\replaced{} {As derived in the Supplementary Section S3, the expected residual as a function of thickness is given by:}
\begin{equation}
R_1(t) \;\approx\; \sqrt{\frac{2}{\pi}}\,\frac{t\,\sigma_\lambda}{2\,\bar\lambda^2}, \quad R_1(t) \propto t
\label{eq:R_final}
\end{equation}
 
so that the slope of the residual is  
\begin{equation}
\frac{\mathrm{d}R_1}{\mathrm{d}t} \;=\; \sqrt{\frac{2}{\pi}}\,\frac{\sigma_\lambda}{2\,\bar{\lambda}^2}.
\label{eq:dRdt}
\end{equation}
$R_1$ is therefore expected to increase linearly with thickness, governed by both the mean absorption length \(\bar{\lambda}\) and its relative spread \(\sigma_\lambda\). A uniform absorptive attenuation, such as that predicted by the two-beam model, is therefore expected to have no impact on residuals, in contrast to the anomalous absorption observed in many-beam simulations.

%Because simulated and experimental intensities generally possess unrelated absolute scales, refinements introduce a global scale factor that places both on a common intensity scale. The refinement residual therefore depends entirely on the relative strengths of reflections. 

\begin{figure}[H]    
    
    \includegraphics[width=1\linewidth]{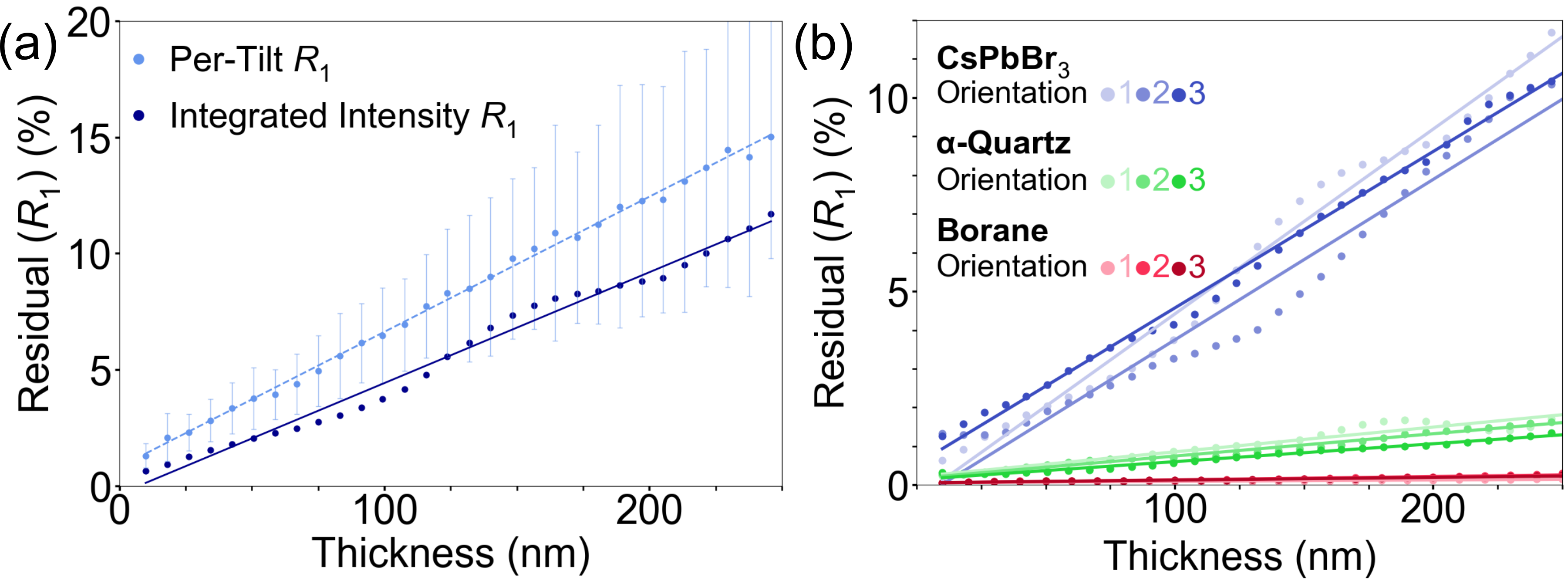}
    \caption{Thickness dependence of residual error \(R_1\) (\%), between simulated $I_\mathrm{abs}$ and $I_{\mathrm{no\,abs}}$. (a) Integrated intensities (dark blue) compared with per-tilt intensities (light blue) for \(\mathrm{CsPbBr_3}\) at \([uvw]=[-0.43,  1.00,  0.93]\); light blue points are \(R_1\) averages across $\pm1.5^\circ$ tilt-series with error bars showing standard deviation around the mean. Best-fit slopes were determined, yielding 
\(0.048~\%~\text{nm}^{-1}\) and \(0.058~\%~\text{nm}^{-1}\) for the integrated intensities and per-tilt values, respectively.
 (b) Integrated-intensity \(R_1\) curves for CsPbBr\(_3\), $\alpha$-quartz, and borane, each shown with three representative orientations (best-fit slopes and \([uvw]\) indices given in Supplementary Table S3.)}
\label{fig:materials}
\end{figure}

To test this, Eq.~\ref{eq:rabs} was evaluated across thickness for  CsPbBr\(_3\) at \([uvw]=[-0.43,  1.00,  0.93]\), with the results shown in Fig.~\ref{fig:materials}(a).

Although the comparison involves only simulated intensities, the calculation was restricted to reflections classified as experimentally observed in the original dataset of \cite{Suresh2024}, with \(I^{\mathrm{exp}}_i/\sigma_i > 3\), \replaced{such that the predicted residuals correspond to those expected for a real experimental refinement.}{  to ensure a consistent and fair comparison with experiment.} \added{The corresponding dataset and crystallographic parameters are summarised in Table~\ref{tab:material_parameters}.}

In Fig.~\ref{fig:materials}(a), the dark blue curve is calculated from integrated intensities, obtained by summing 60 tilts across the full \(\pm 1.5^\circ\) tilt series, while residuals at individual tilts (non-integrated intensities) are shown in light blue. The comparison demonstrates reduced anomalous absorption achieved by integration: although individual tilts can show strong deviations, these largely cancel when averaged.

At low thicknesses the predicted effect of absorption is minimal, with \(R_1\) remaining below 5\% for thicknesses under 100~nm. Using the parameters \(\bar{\lambda}=93.2~\text{nm}\) and \(\sigma_{\lambda}=10.1~\text{nm}\), the predicted slope from Eq.~\eqref{eq:dRdt} is 
$\frac{\mathrm{d}R_1}{\mathrm{d}t} = 0.046~\%\!~\text{nm}^{-1}$, in close agreement with the best-fit slope of \(0.048~\%\!~\text{nm}^{-1}\) shown in Fig.~\ref{fig:materials}(a).

This residual can be interpreted as a nominal attainable lower bound on an elastic-only dynamical refinement. To compare this predicted lower bound with real refinements, we consider the elastic-only refinements of the same CsPbBr$_3$ dataset reported by Suresh \textit{et al.} and re-analysed by Malik \textit{et al.}~\cite{malik2025hybrid}. Suresh \textit{et al.} reported an overall $R_{\mathrm{obs}} = 5.3\%$, while Malik \textit{et al.} yielded a value of $R_{\mathrm{obs}} = 6.5\%$. Because the present study employs the refinement framework and simulation pipeline of Malik \textit{et al.}, all comparisons refer specifically to their results, ensuring that parameters are defined and controlled consistently. 

While the true specimen thickness is not known, the hybrid physics-machine learning framework used in this study and introduced in \cite{malik2025hybrid} extracts an orientation-dependent thickness distribution directly from the diffraction data through joint refinement of structural and experimental parameters. For the orientation shown in Fig.~\ref{fig:materials}(a), this corresponds to a thickness of $t \approx 50 $ nm, for which the predicted effect of neglecting absorption is around 2\%. This comparison shows that the low refinement residuals reported are consistent with the predicted effect of absorption at this thickness, indicating that only a minor improvement would be expected from incorporating absorption in the refinement. 

While the preceding comparison concerns only a single orientation, absorption and resulting residuals are expected to vary with orientation. To quantify this effect, \(R_1(t)\) between absorptive and purely elastic simulations was computed for a range of orientations and materials typical of 3D~ED experiments. Figure~\ref{fig:materials}(b) compares these results for three materials taken from recently reported continuous-rotation 3D~ED datasets: \(\mathrm{CsPbBr_3}\)~\cite{Suresh2024}, $\alpha$-quartz~\cite{klar_accurate_2023}, and octadecaborane (hereafter referred to as borane) ~\cite{Suresh2024}. 

In each case, the first three experimentally observed orientations were simulated, with Eq.~\ref{eq:rabs} evaluated only for reflections with \(I^{\mathrm{exp}}_i/\sigma_i > 3\). These materials were selected to capture a wide range of structural and physical characteristics: \(\mathrm{CsPbBr_3}\) is a halide perovskite with high atomic number, quartz is an inorganic oxide with moderate atomic number, and borane is an inorganic molecular crystal composed of light elements. \added{Additionally, \(\mathrm{CsPbBr_3}\) (\(Pbnm\)) and borane (\(Pccn\)) are centrosymmetric, whereas \(\alpha\)-quartz (\(P3_221\)) is non-centrosymmetric.} Their crystallographic parameters are listed in Table~\ref{tab:material_parameters}.

\begin{table}[H]
\centering
\small
\caption{Crystallographic parameters for the three materials used in this study. \added{For full experimental acquisition and data reduction details, see original references.} }
\label{tab:material_parameters}
\begin{tabular}{l p{3.3cm} p{3.3cm} p{3.5cm}}
\hline
\textbf{Parameters} 
& \textbf{CsPbBr$_3$} 
& \textbf{$\alpha$-quartz (SiO$_2$)} 
& \textbf{Borane (B$_{18}$H$_{22}$)} \\

Source dataset 
& \cite{Suresh2024}
& \cite{klar_accurate_2023}
&  \cite{Suresh2024} \\

Space group 
& $Pbnm$ 
& $P3_221$ 
& $Pccn$ \\

{Unit cell $(a, b, c)$ (\AA) }
&8.119, 8.359, 11.759
&4.923, 4.923, 5.400
&10.770, 11.990, 10.736\\

Angles $(\alpha, \beta, \gamma) (^\circ) $
& 90, 90, 90 
& 90, 90, 120 
& 90, 90, 90 \\

Volume (\AA$^3$)
& 798.1
& 113.3
& 1386.9 \\ 

%Observation temperature (K)
%& 150 
%& 293 
%$& 100 \\

%Debye temperature $\Theta_D$ (K) 
%& 102~\cite{evarestov_first-principles_2020}
%& 254~\cite{lord_calculation_1957}
%& 135~\cite{andersson_low-temperature_1993} \\
\hline
\end{tabular}
\end{table}

From Fig.~\ref{fig:materials}(b), clear material-dependent trends emerge. For all three materials, the first three orientations lie within a narrow mean-thickness range, giving values of approximately 50 nm for CsPbBr$_3$, 85 nm for $\alpha$-quartz, and 170 nm for borane, as determined from the orientation-dependent thickness profile obtained in the elastic-only refinements (See Supplementary Fig.~S8). At these thicknesses, absorption contributes ca. 2\% for CsPbBr$_3$, 0.6\% for $\alpha$-quartz, and 0.2\% for borane. 

As shown earlier, in the two-beam limit, inclusion of absorption produces almost no per-reflection variation, with intensities attenuated uniformly. Under many-beam conditions, however, absorption alters the inter-beam coupling, changing how intensity is redistributed with thickness. The resulting deviation from the elastic-only solution therefore reflects the degree of many-beam interaction. In \(\mathrm{CsPbBr_3}\), the heavy constituent atoms lead to the simultaneous excitation of many strong beams. In contrast, $\alpha$-quartz and borane exhibit much weaker many-beam character at these thicknesses. Their lower atomic numbers result in fewer simultaneously excited strong reflections, so scattering is dominated by quasi-two-beam interactions. In such regimes, the inclusion of absorption does not result in significant redistribution of diffracted intensity. \added{The behaviour observed for quartz further suggests that the additional absorptive effects associated with non-centrosymmetric crystals remain comparatively weak under the present conditions.}

\subsection{Orientation Dependence of Absorption Effects}

In addition to the strong material dependence, little systematic variation is observed across orientations in Fig. \ref{fig:materials}(b). This may be explained by the geometry of the orientations examined, with each lying away from strongly diffracting zone axes. This behaviour is characteristic of 3D~ED datasets, where randomly orientated crystals are rotated through large angular ranges that remain predominantly off-zone, typically intersecting perhaps only a single major zone axis over the course of the tilt series. At such zone-axis conditions, however, where many-beam effects are strongest, a correspondingly greater influence of absorption is expected. 

\begin{figure}[H]    
    \includegraphics[width=0.9\linewidth]{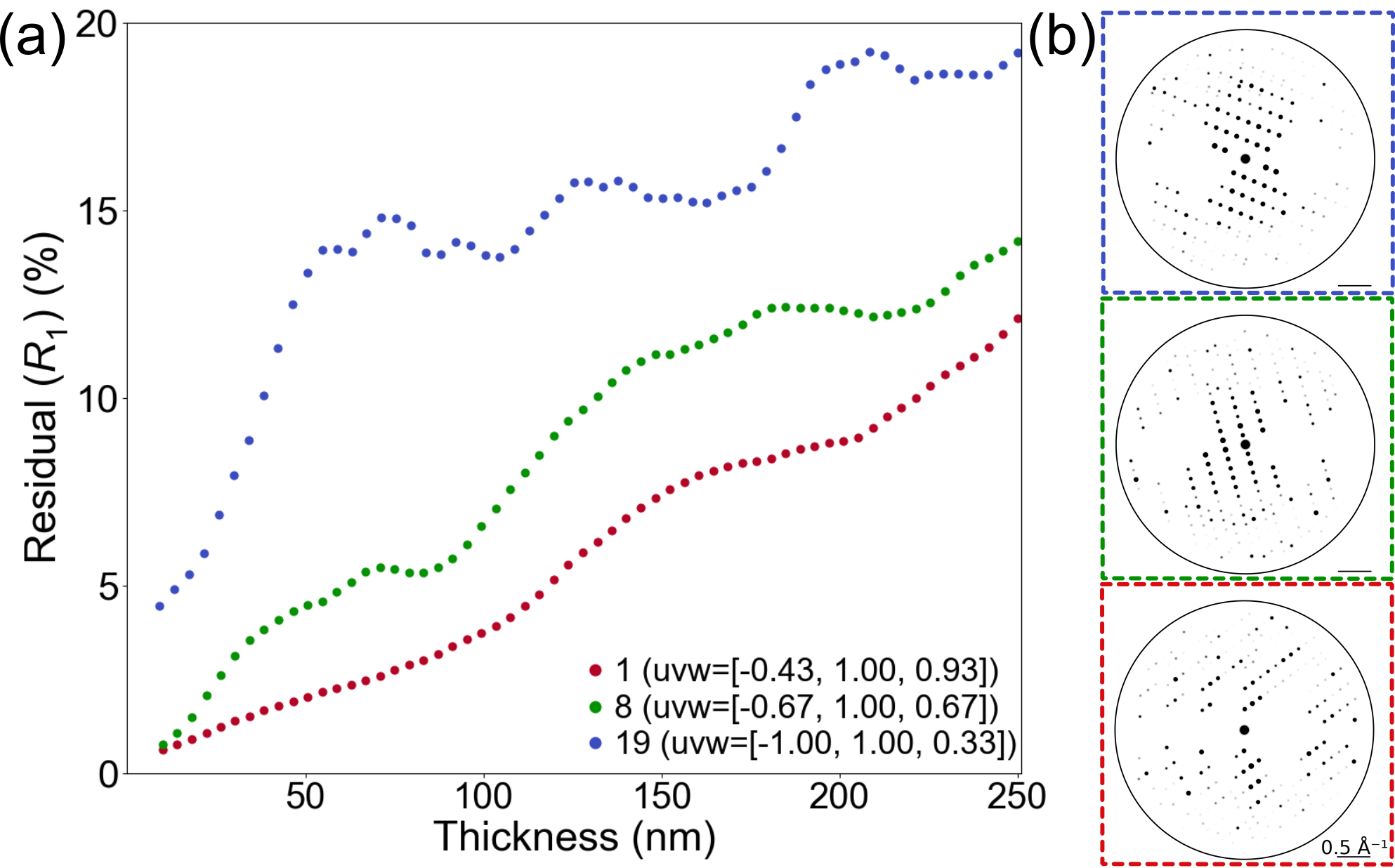}
    \caption{Orientation dependence of absorption effects in $\mathrm{CsPbBr_3}$. (a) Thickness dependence of $R_1$ between simulated  $I_\mathrm{abs}^{\mathrm{int}}$ and $I^{\mathrm{int}}_{\mathrm{no\,abs}}$ for orientations~1, 8, and 19. (b) Corresponding simulated diffraction patterns, each enclosed by a dashed box in the colour of its respective curve in (a).}
\label{fig:zone}
\end{figure}

To examine this behaviour explicitly, three representative orientations from the CsPbBr$_3$ dataset were selected to span off-zone, intermediate, and near-zone regimes. While several orientations in the dataset fall into each category, orientations 1, 8, and 19 were chosen for illustration, with the corresponding simulated diffraction patterns shown in Fig.~\ref{fig:zone}(b). Orientation~1 ($[uvw]$=[-0.43, 1.00, 0.93]) serves as the representative off-zone reference, orientation~8 ($[uvw]$=[-0.67, 1.00, 0.67]) corresponds to a direction near [$\bar{2}32$], and orientation~19 ($[uvw]$=[-1.00, 1.00, 0.33]) lies close to [$\bar{3}31$].

In Fig.~\ref{fig:zone}(a), the variation of \(R_1(t)\) with crystal orientation is shown, highlighting the strengthening of many-beam coupling and the associated enhancement of anomalous absorption effects as the crystal is tilted toward a strongly diffracting zone axis. Orientation~1 (seen previously in Fig. \ref{fig:materials} (a,b)) is representative of most orientations in the dataset, with an approximately linear thickness dependence in $R_1$. Orientation~8 displays a markedly steeper trend, indicating stronger many-beam coupling, while orientation~19 shows the strongest and most nonlinear response, rising to nearly 13\% by 50\,nm. Such a large residual indicates that an elastic-only refinement would perform poorly for this orientation, with a correspondingly large improvement expected if absorption is included.

Further insight into this behaviour is provided in Supplementary Fig. S6, in which \(I^{\mathrm{int}}_{\mathrm{abs}}/I^{\mathrm{int}}_{\mathrm{no\,abs}}\) is plotted as a function of thickness for the corresponding curves. These simulations show that the decaying-exponential attenuation underlying the linear \(R_1(t)\) trend no longer holds as the crystal approaches a zone axis. %The corresponding mean absorption lengths $\bar{\lambda}$ decrease systematically from 93.8\,nm for orientation~1 to 89.4\,nm for orientation~8 and 85.2\,nm for orientation~19, with $\sigma_{\lambda}$ values of 8.4, 11.7, and 22.1\,nm, respectively, reflecting the progressively stronger influence of anomalous absorption under many-beam conditions.

Comparable behaviour has been reported in recent dynamical refinements, where orientations near zone axes exhibit anomalously high residuals, previously ascribed to crystal imperfections and excessive dynamical effects \cite{Olech2024}. In that study, fifteen frames corresponding to these orientations were omitted from the refinement, representing a significant loss of usable data. The present results indicate that exclusion of such frames may be unwarranted, as the elevated residuals are largely attributable to previously unmodelled absorption effects. 

\subsection{Elastic and Absorptive Refinements of Experimental 3D-ED Data}

Although the preceding simulations indicate that the inclusion of absorption will significantly alter simulated intensities, it remains unclear whether these differences will yield measurable benefits in refinement accuracy or in the recovered structure. To answer this question, full dynamical refinements were carried out for CsPbBr$_3$, $\alpha$-quartz and borane, with and without absorption. 

For each material, refinements followed the complete workflow described in \cite{malik2025hybrid}. %This included an initial orientation refinement, producing self-consistent sets of refined orientations for the elastic and absorptive models. 
This included orientation-dependent thickness refinements. The corresponding refined thickness–tilt curves for all three materials are shown in Supplementary Fig.~S8; overall the absorptive and elastic models give similar trends, but notable deviations occur for specific orientations (e.g., CsPbBr$_3$ $\sim$55 vs 48~nm; borane $\sim$200 vs 170~nm).

\replaced{}{The corresponding refined thickness–tilt curves for all three materials are shown in Supplementary Fig.~S8}. The resulting refinement statistics are summarised in Table~\ref{tab:refinement_residuals}, with refinement parameters provided in Supplementary Table S4.

\begin{table}[H]
\centering
\small
\caption{Refinement residuals for CsPbBr$_3$, $\alpha$-quartz, and borane.}
\label{tab:refinement_residuals}
\renewcommand{\arraystretch}{1.25}
\setlength{\tabcolsep}{10pt}

\begin{tabular}{lccc}
\hline
\textbf{Residual (\%)}& \textbf{CsPbBr$_3$} 
& \textbf{$\alpha$-quartz} 
& \textbf{Borane} \\
\hline
Elastic $R_{\mathrm{obs}}/wR_{\mathrm{all}}$ 
& 6.40 / 6.73& 4.14 / 3.84& 9.54 / 8.56\\
Absorptive $R_{\mathrm{obs}}/wR_{\mathrm{all}}$
& 5.26 / 5.31& 4.00 / 3.66& 9.48 / 8.51\\
\hline
\end{tabular}
\end{table}

Across all three materials, the inclusion of absorption produces only modest changes in the refinement residuals. CsPbBr$_3$ shows a clear improvement when absorption is included, while the refinement residuals of $\alpha$-quartz and borane remain effectively unchanged. 

While the \replaced{earlier}{ eariler} analysis predicted a reduction of $R_{\mathrm{obs}}$ for CsPbBr$_3$ from approximately 6.4$\%$ to $4.5\%$, the refinement only reaches 5.3 $\%$. This outcome likely reflects both the simplifying assumptions of the absorptive model, including its isotropic treatment of thermal motion and unmodelled phonon correlations, and the broader set of factors that contribute to the observed residual, such as plasmon and core-loss ionisation, beam damage, and crystal imperfections. These factors obscure the specific contribution of absorption, and the observed decrease in $R_{\mathrm{obs}}$ is therefore consistent with the expected behaviour under experimental conditions.

For quartz and borane, the predicted absorption effects were modest, and the refinements exhibit correspondingly small improvements. For both materials, absorption therefore plays no significant role in the refinement, and the remaining analysis focuses on CsPbBr$_3$, with the full quartz and borane results reported in the Supplementary Material. 

\begin{figure}[H]    
    \centering
    \includegraphics[width=0.9\linewidth]{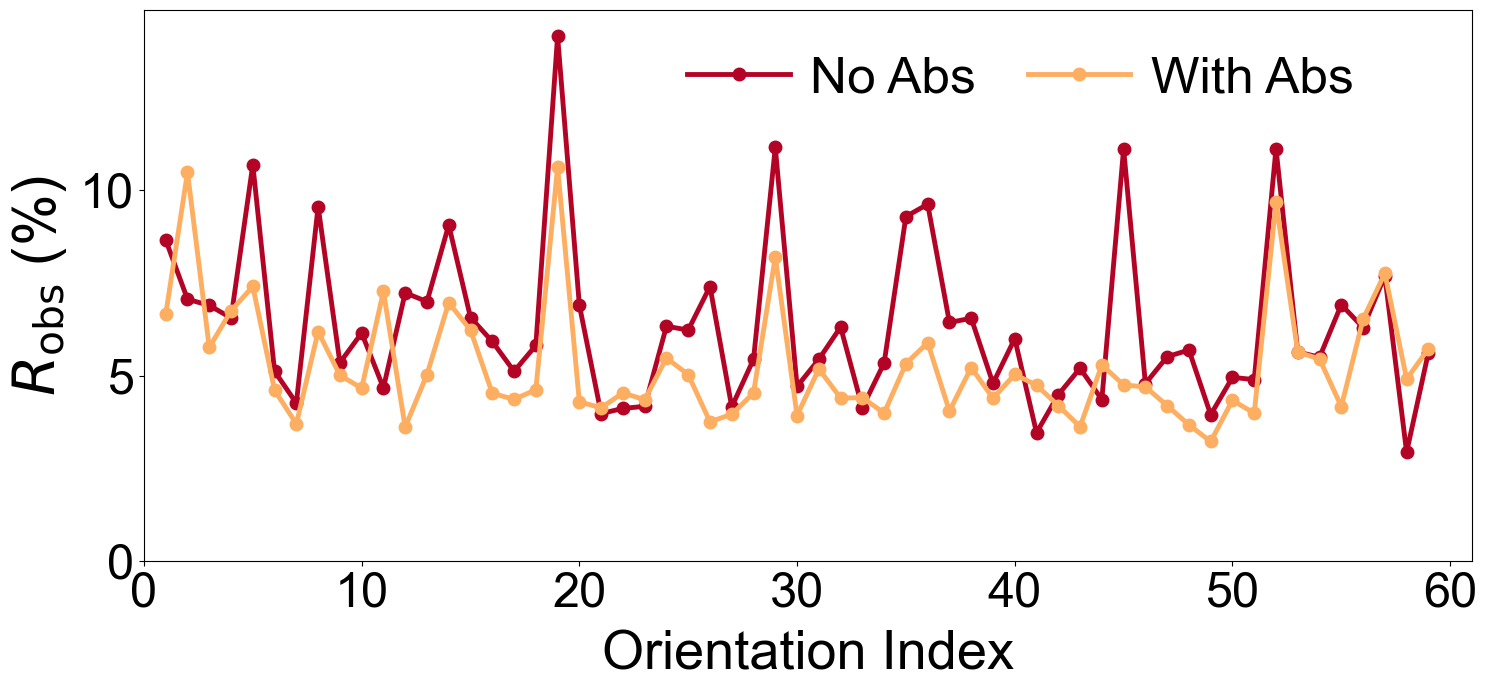}
    \caption{Results of dynamical refinements performed with and without the inclusion of absorption for CsPbBr$_3$. Residuals $R_{obs}$ (\%) are compared across rotation indices for refinements carried out on identical datasets and parameters. Supplementary Fig. S7 shows the corresponding $wR_\mathrm{all}$ results for CsPbBr$_3$ and the full orientation-dependent residuals for $\alpha$-quartz and borane.}

\label{fig:refinement}
\end{figure}

Beyond the overall improvement in residuals observed for CsPbBr$_3$,  the orientation-dependent residuals (Fig.~\ref{fig:refinement}) show several pronounced outliers (orientations~19, 29, 46, 53). As discussed in Section 3.2, these orientations lie near strongly diffracting zone axes, where many-beam coupling is stronger and absorption has a larger influence. 

Although the residuals of orientations~19, 29, 53 decrease markedly following the inclusion of absorption, they remain high outliers. This behaviour indicates that the present absorptive model corrects only part of the deviation and that additional zone-axis effects, possibly including plasmon and core-loss ionisation contributions or more complex dynamical interactions, remain unaccounted for. While such orientations might warrant exclusion in an elastic-only refinement, the absorptive model provides a sufficient basis for their retention. 

Additionally, several orientations exhibit worse residuals under the absorptive model. While it is not clear why this occurs, \replaced{this may be attributable to experimental uncertainty together with the approximate treatment of absorption employed here.} { it likely reflects the non-convex nature of the refinement, in which the search trajectory may settle into a suboptimal local minimum.}

To further assess the consequences of including absorption on the refined structure, Supplementary Table S5 compares the atomic parameters obtained with and without the absorptive potential for CsPbBr$_3$. The inclusion of absorption produces only minor shifts in both fractional coordinates and anisotropic displacement parameters. The largest observed positional difference is approximately $0.001$ in fractional coordinates, with corresponding changes in anisotropic displacement parameters below $0.003~\text{\AA}^2$, both within normal refinement precision \cite{Brazda2019}, confirming that the refined structure remains stable under the modified scattering model. This indicates that even for a material with large mean atomic number \(\langle Z \rangle\) of 48.4, for thicknesses below 50 nm, absorption has little practical impact on the recovered structure. Consequently, omitting absorption from dynamical refinement simulations is expected to bias thermal and structural parameters only in high-$Z$ materials at thicknesses approaching $\xi_g$.

\section{Conclusion}

This work presents what is, to our knowledge, the first implementation of absorption in 3D~ED dynamical refinement, establishing its influence on experimental residuals and refined structures. Though central to CBED contrast, absorption has a much weaker influence on the relative intensities measured in 3D ED. This discrepancy arises from integration, which averages orientation-specific anomalous absorption.

The underlying theory is examined, beginning from the two-beam approximation, yielding a closed form expression for integrated intensities in the presence of absorption \added{for centrosymmetric crystals}. This shows that for \(t \ll \xi_{g}\) reflections experience a uniform exponential decay with thickness, with a decay rate set by the mean absorptive potential \(U_{0}'\). In the many-beam limit, dynamical refinements that neglect absorption incur a systematic, thickness-dependent residual. The residual \(R_1(t)\) grows approximately linearly with thickness, determined by the mean absorption length \(\bar{\lambda}\) and its spread \(\sigma_{\lambda}\), and \replaced{with this effect becoming more severe near zone axes}{ diverging near zone axes}. This behaviour explains a longstanding observation that orientations near zone axes exhibit anomalously high residuals in elastic-only dynamical refinements and provides a framework for incorporating these data rather than discarding them.

%This was further shown to break down as \(t\) approaches \(\xi_{g}\), where reflections begin to deviate from the uniform behaviour.

Comparative refinements of CsPbBr$_3$ performed with and without absorption show a clear improvement in fit when absorption is included, reducing the refinement residual \(R_{\mathrm{obs}}\) from 6.4\,\% to 5.3\,\%. In contrast, refinements of $\alpha$-quartz and borane under identical conditions show little improvement in \(R_{\mathrm{obs}}\), consistent with expectations for lower-Z materials. 

Absorption can therefore safely be neglected for routine refinements except in high-Z materials for thicknesses approaching $\xi_g$. Although not addressed in the present analysis, temperature is also expected to play a significant role, with materials observed at $T \ge \Theta_D$ likely to exhibit strong absorptive effects. 

Additionally, the influence of omitting absorption in the determination of finer structural information such as bonding charge density remains unclear. It is worth noting that deviations away from neutral atom scattering factors will affect low-order reflections disproportionately. These have, in general, the largest $U_g, U_g'$ and the effects of absorption will likely be appreciable. 

\added{Similarly, although the additional absorptive effects associated with non-centrosymmetric crystals were shown here to be comparatively small, these may still provide useful sensitivity to chirality and absolute structure determination through the breaking of Friedel equivalence.}

\section*{Acknowledgements}
The authors thank Lukas Palatinus and his co-workers for access to their 3D ED data used in this work. The authors would further like to thank Lukas Palatinus, Ondrej Krivanek, Colin Humphreys, Mike Treacy, Petr Vacek, Hannah Cole and Simon Fairclough for helpful discussions.

\section*{Funding}
BC acknowledges funding from Queens' College Cambridge and the Stamps Scholars Program. TASD acknowledges the support of a Schmidt Science Fellowship. SM acknowledges funding from EPSRC Centre for Doctoral Training in Autonomous Intelligent Machines and Systems (Grant No: EP/S024050/1). PAM acknowledges funding from the Engineering and Physical Sciences Research Council (Nos. EP/W522120/1 and EP/R008779/1).

\section*{Conflicts of Interest}
The authors disclose no conflicts of interest.

\section*{Data Availability}
All data, analyses, and code used to produce the results are available in the GitHub repository bcolmey/role-of-absorption-in-3DED-dynamical-refinement.

\printbibliography

\end{document}